\documentclass{optica-article}

\journal{opticajournal} 

\articletype{Research Article}
\usepackage{lineno}

\usepackage{amsmath, bm, upgreek, overpic, multirow, graphicx, xcolor}

\DeclareMathAlphabet{\mathbbold}{U}{bbold}{m}{n}
\renewcommand{\a}{\alpha}
\renewcommand{\b}{\beta}

\renewcommand{\d}{\delta}
\newcommand{\e}{\varepsilon}

\renewcommand{\k}{\kappa}

\newcommand{\m}{\mu}

\newcommand{\x}{\xi}
\newcommand{\p}{\pi}
\renewcommand{\r}{\rho}
\newcommand{\s}{\sigma}
\renewcommand{\t}{\tau}
\renewcommand{\u}{\upsilon}
\newcommand{\ph}{\varphi}
\renewcommand{\c}{\chi}

\renewcommand{\o}{\omega}

\newcommand{\Scal}{\mathcal S}

\newcommand{\vc}[1]{\boldsymbol{\mathrm{#1}}}
\newcommand{\ts}[1]{\hat{\mathrm{#1}}}

\newcommand{\uv}[1]{\hat{\boldsymbol{\mathrm e}}_{#1}}
\newcommand{\tsone}{\hat{\mathbbold{1}}}

\newcommand{\bra}[1]{\left< #1 \right|}
\newcommand{\ket}[1]{\left| #1 \right>}

\renewcommand{\div}{\nabla\cdot}
\newcommand{\rot}{\nabla\times}
\renewcommand{\Im}{\mathrm{Im}}
\renewcommand{\Re}{\mathrm{Re}}

\newcommand{\lrp}[1]{\left(#1\right)}
\newcommand{\lrb}[1]{\left[#1\right]}
\newcommand{\lrc}[1]{\left\{#1\right\}}
\newcommand{\lrv}[1]{\left|#1\right|}
\newcommand{\lra}[1]{\left< #1 \right>}

\begin{document}
\title{Optical Spintronics: Towards Optical Communication Without Energy Transfer}

\author{Ilya Deriy\authormark{1,2},
Danil Kornovan\authormark{2},
Mihail Petrov\authormark{2}, 
Andrey Bogdanov\authormark{1,2,*}}

\address{
\authormark{1}Qingdao Innovation and Development Center, Harbin Engineering University, Sansha Rd. 1777, Qingdao 266000, Shandong, China 
\\
\authormark{2}Department of Physics and Engineering, ITMO University, St. Petersburg 197101, Russia
}
\email{\authormark{*}a.bogdanov@hrbeu.edu.cn}

\begin{abstract*}
We introduce the concept of optical spin current, wherein optical spin angular momentum is transferred by an electromagnetic field without accompanying energy transfer. This phenomenon is analogous to electron spin currents, where spin is decoupled from charge flow. Building on this principle, we propose an optical spin diode and an optical spin circulator—devices that enable unidirectional propagation of spin currents while maintaining bidirectional energy flow, thus, preserving reciprocity. Furthermore, we demonstrate asymmetric spin transfer between two quantum dots mediated by the optical spin diode, highlighting the potential for novel optical spintronic functionalities. These findings lay the foundation for devices that use optical spin transfer, opening new avenues for advancements in optical spintronics.
\end{abstract*}

\section{Introduction}\label{section label}
{\it Photonics} has emerged as a promising successor to electronics, offering numerous advantages such as lower material losses, higher bandwidth, and the potential for faster data transmission speeds \cite{li2024exploring, mcmahon2023physics}. Unlike traditional electronics, which rely on the flow of electrons, photonics benefits from the unique properties of light—particularly the high speed and effectively massless nature of photons. As a result, photonic devices tend to generate less heat, significantly improving their operational frequency and energy efficiency compared to their electronic counterparts \cite{chanana2022ultra, dong2021ultra, blundell2025ultracompact}.

{\it Spintronics} is another rapidly developing field positioned as a successor to conventional electronics. It exploits the intrinsic spin of electrons, rather than their charge, for information storage and processing \cite{bader2010spintronics, pulizzi2012spintronics, vzutic2004spintronics}. This shift away from charge-based systems holds great promise for reducing energy consumption. By using spin as an additional degree of freedom, spintronics extends the functionality of electronic devices and enables new physical mechanisms for more energy-efficient data storage and logic operations \cite{qin2023spintronic, yang2022two}.

Much like electrons, photons also possess angular momentum, which plays a critical role in their interactions with matter. The angular momentum has two components: spin angular momentum (SAM) and orbital angular momentum (OAM) \cite{bliokh2013dual, bliokh2015spin, bliokh2015transverse}. SAM is associated with the photon's polarization, while OAM arises from the spatial distribution of the photon's phase front. These two forms of angular momentum provide additional control over the propagation and interaction of light \cite{liu2022photonic, zhou2018broadband}. The ability to manipulate both SAM and OAM opens up new avenues for information encoding and transfer, making angular momentum a valuable resource in photonics.
Utilization of angular momentum of photons allows for multistate information encoding and high-capacity multiplexing, which has proven to be useful in optical communication \cite{marrucci2006optical, marrucci2011spin} and quantum computing~\cite{nagali2010experimental, erhard2018twisted}.
Independence of SAM and OAM on the intensity of light allows for nanoscale imaging and enhanced spectroscopy even in the low-intensity regions leading to optical trapping, imaging, and sensing beyond diffraction limit~\cite{chen2021engineering, paul2022simultaneous}.
Moreover, rich nature of spin-orbital light-matter interactions makes angular momentum of photons useful in many emergent applications like reconfigurable optical devices~\cite{mao2022active}, twisted photonics systems~\cite{vyatkin2025emergent}, and integrated photonic chips~\cite{paul2022simultaneous}.

Although photons, like electrons, carry spin angular momentum, the idea of using optical spin as the main information-carrying quantity for information transfer is still in its early stages compared to the well-established field of classical spintronics.

Reduced energy dissipation and faster spin states dynamics compared to charge states allows for the creation of faster, more energy efficient non-volatile spin-based memory~\cite{puebla2020spintronic, scivetti2021combined, joshi2016spintronics, yang2022two}.
Moreover, spin-based memories also allow for multiple bits per cells, enabling ultrahigh-density storage~\cite{puebla2020spintronic, joshi2016spintronics}.
In addition, spin-based devices have shown themselves to be more resistant to temperature and cosmic radiation, enhancing the device longevity under harsh conditions~\cite{vzutic2004spintronics}.
Additionally, spintronic devices enable in-memory computing, which makes them very attractive in the era of artificial intelligence~\cite{chen2023spintronic}.
While spintronics has led to a wide range of practical applications by exploiting the spin of electrons for data storage and processing, the potential of photon spin remains largely untapped. 

In this work, we draw an analogy to spintronics and propose a novel direction for photonic technologies -- {\it optical spintronics} (see Fig.~\ref{fig:main_idea}a). Just as spintronics exploits the electron’s spin to extend the capabilities of electronics, optical spintronics leverages the spin angular momentum (SAM) of photons for information transfer and processing. We introduce the concept of an optical spin current, defined as the transfer of SAM by light, and demonstrate that specific electromagnetic field configurations can support a nonzero SAM flux without any accompanying energy transfer similarly to electron spin currents, which convey spin without charge flow. This concept offers a new paradigm for low-energy, efficient optical computing. We also show that chiral media can serve as {\it optical isolator} for spin current, enabling unidirectional propagation of spin currents while preserving reciprocal energy flow. Based on this principle, we propose the design of optical spin diode and circulator as basic components of optical spintronic circuits. Finally, we numerically demonstrate the unidirectional transfer of optical spin current between two quantum dots mediated by a chiral environment, illustrating the practical feasibility of spin-based information transport.

\section{Optical Spin Current}
\label{sec:optical_spin_current}
The angular momentum of the electromagnetic field can be expressed as the sum of the orbital angular momentum, which is associated with the spatial distribution of the light beam, and the SAM, which is associated with the polarization properties of light~\cite{bliokh2017optical}.
As electromagnetic energy and linear momentum, SAM also satisfies continuity equation (see Appendix, Sec.~\ref{apdx:SAM_continuity} or Ref.~\cite{tanaka2020continuity})
\begin{equation}
    \div \ts J_\mathrm{S} + \dfrac{\partial}{\partial t}\vc S = 0,
    \label{eq:continuity_equation}
\end{equation}
where $\vc S$ is the optical SAM density, and $\ts J_\mathrm{S}$ is the optical SAM density flux, with $\ts J_{\mathrm{S}_{ij}}$ showing the flux of the $j$-th component of SAM $S_j$ in $i$-th direction.
For monochromatic waves in vacuum, time-averaged SAM density and SAM flux density read as (see Appendix, Sec.~\ref{apdx:wavepacket} or Ref.~\cite{bliokh2017optical})
\begin{gather}
    \lra{\vc S} = \dfrac{1}{16 \p \omega }\Im\lrb{\vc E^* \times \vc E + \vc H^* \times \vc H},
    \\
    \langle \ts J_\mathrm{S} \rangle = \dfrac{c}{16 \pi \omega}\Im\lrb{\tsone \lrp{\vc E\cdot \vc H^*} - \vc E\otimes \vc H^* - \vc H^*\otimes \vc E},
\end{gather}
where $\omega$ is the frequency of the wave $k = \o/c$ is the wavevector, and $c$ is the speed of light.
We will refer to SAM flux as \textit{optical spin current}.
It is important to note that for clarity and ease of understanding of the underlying physical processes related to the optical spin currents, we will mostly consider only one component of the optical spin current, the one which corresponds to the flow in the propagation direction of the wave of the SAM projection on the propagation direction of the wave.

\begin{figure}[t]
	\center
	\includegraphics[width=0.8\linewidth]{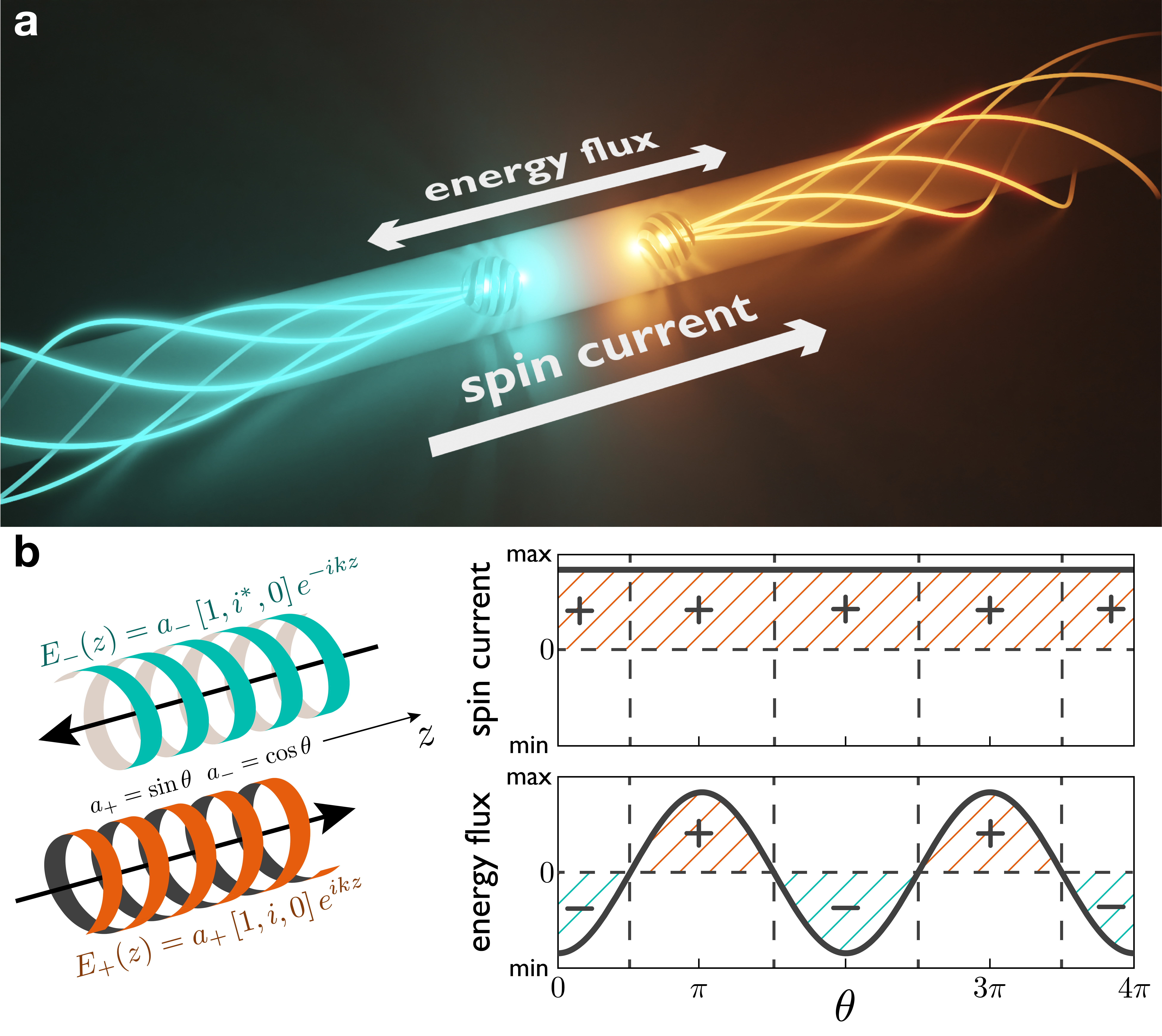}
	\caption{Optical Spintronics. \textbf a - An artistic view at one of the possible application of optical spintronics, an optical spin diode. \textbf b – Electromagnetic field configuration with zero energy flux, but nonzero optical spin current.}
	\label{fig:main_idea}
\end{figure}

For a plane wave in vacuum, the optical spin current density is equal to the outer product of SAM density and the group velocity $\vc v_\mathrm{g}$ (see Appendix, Sec.~\ref{apdx:wavepacket})
\begin{equation}
    \langle \ts J_\mathrm{S} \rangle = \dfrac{1}{2}\lra{\vc S}\otimes \vc v_\mathrm{g}.
    \label{eq:spin_plane}
\end{equation}
This expression allows us to draw analogies between photon spin and electron currents.
Electromagnetic waves with the same sign of the SAM and the group velocity have a positive spin current, and vice versa.
This leads to the following observation – superposition of two waves propagating in different directions can have a non-zero spin current, but, at the same time, zero Poynting vector. This happens when in each of the waves the sign of SAM and group velocity are same (see Fig.~\ref{fig:main_idea}b).
Indeed, for the following field configuration 
\begin{equation}
    \vc E\lrp{\vc r} = a_+\uv{0} \: e^{i k z} + a_-\uv{0}^*\: e^{- i k z},
    \label{eq:field_spin_current}
\end{equation}
where $\vc e_0 = \lrp{\uv{x} + i \uv{y}}/\sqrt{2}$, and $a_\pm$ are complex amplitudes, only nonzero components of energy flux (Poynting vector) and spin current density are
\begin{equation}
\begin{gathered}
    \lra{\vc P}_z = \dfrac{c}{8\pi}\lrp{\lrv{a_+}^2 - \lrv{a_-}^2 },
    \\
    \langle \ts J_\mathrm{S} \rangle_{zz} = \dfrac{c}{16 \pi \omega}\lrp{\lrv{a_+}^2 + \lrv{a_-}^2 }.
\end{gathered}
\end{equation}
When the amplitudes of forward and backward propagating waves are equal, they form a standing wave, however spin current is nonzero (see Fig.~\ref{fig:main_idea}b).

\section{Nonreciprocal Optical Spintronic Devices}
\label{sec:FiguresTables}
\begin{figure*}[t]
	\center
	\includegraphics[width=\linewidth]{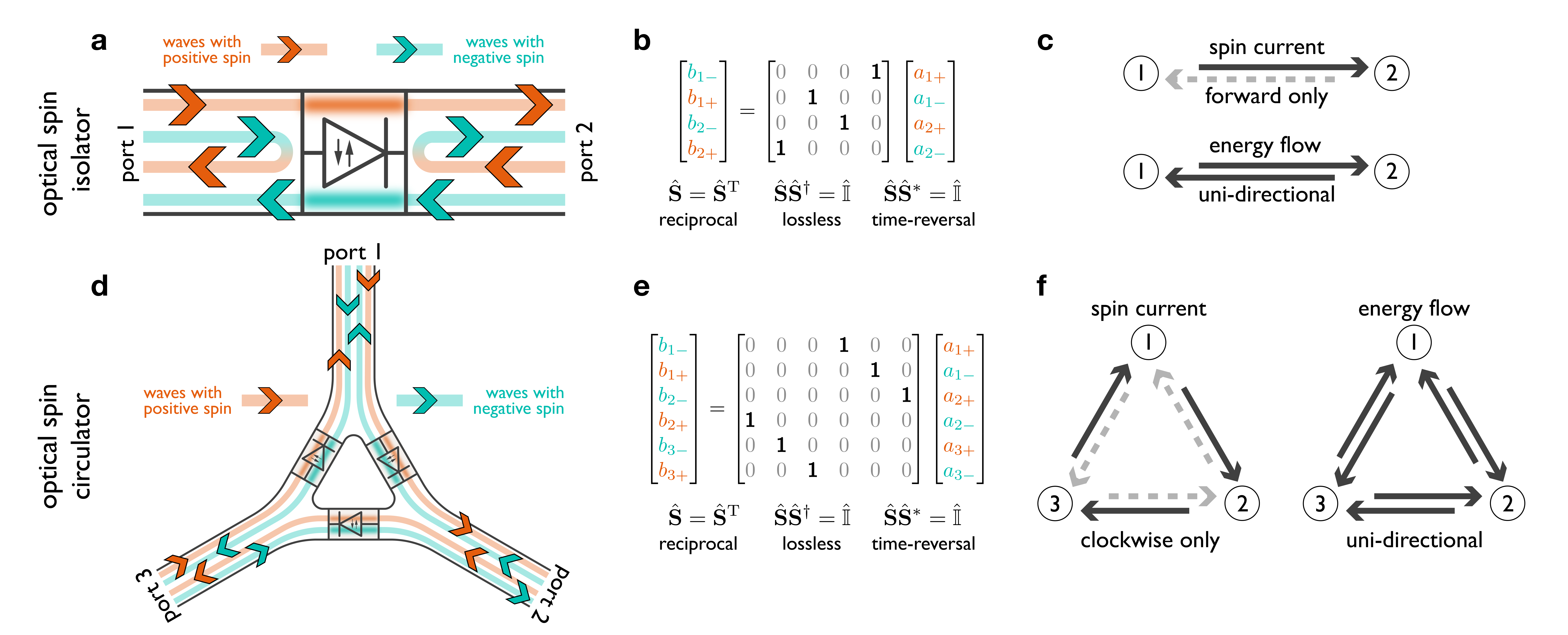}
	\caption{Nonreciprocal optical spin devices.
    \textbf a, \textbf d – schematic geometry, \textbf b, e – scattering matrices, \textbf c, \textbf f – spin currents and energy flow in optical spin isolator and circulator respectively.}
	\label{fig:diode_ideal}
\end{figure*}

It follows from Eq.~\eqref{eq:spin_plane} that in the free space, there are no restrictions on the directions of propagation of optical spin currents.
However, such restrictions can be imposed by using various optical devices.
The simplest example of such a device is an optical spin isolator – a two-port optical network in which optical spin current is able to flow only in one direction.
Consider an abstract two-port network with each port supporting two orthogonal polarizations of electromagnetic waves.
Inputs and outputs of such a network can be presented as plane waves propagating along $z$ direction,
\begin{equation}
    \vc E_i\lrp{\vc r} = \lrp{\uv{+}a_{i+} + \uv{-}a_{i-}} e^{i \b_i k z} + \lrp{\uv{+}b_{i+} + \uv{-}b_{i-}} e^{-i \b_i k z},
    \label{eq:sc_field}
\end{equation}
where $ i = 1$ ($i = 2$) encodes fields to the left (to the right) of the network (see Fig.~\ref{fig:diode_ideal}a), ${\b_1 = 1 = - \b_2}$.
The polarizaion basis $\uv{\pm} = \left(\uv{x} \pm i \uv{y}\right)/\sqrt{2}$ is chosen in such a way that index $\pm$ encodes sign of the $z$ component of SAM $\lra{\vc S}_z$.
Transmission and reflection properties of such a two-port network are fully defined by its scattering matrix, which connects amplitude vectors of incident and scattered waves as follows $\vc b = \ts S \vc a$.
As the simplest theoretical concept of an optical spin isolator, we present the following scattering matrix
\begin{equation}
    \begin{bmatrix}
        b_{1-}
        \\
        b_{1+}
        \\
        b_{2-}
        \\
        b_{2+}
    \end{bmatrix}
    =
    \underbrace{\begin{bmatrix}
        0 & 0 & 0 & 1
        \\
        0 & 1 & 0 & 0
        \\
        0 & 0 & 1 & 0
        \\
        1 & 0 & 0 & 0
    \end{bmatrix}}_{\ts S}
    \begin{bmatrix}
        a_{1+}
        \\
        a_{1-}
        \\
        a_{2+}
        \\
        a_{2-}
    \end{bmatrix}.
    \label{eq:smatrix_ideal}
\end{equation}
In the two-port network with such a scattering matrix, only waves with a positive (negative) SAM will be able to pass through in the positive (negative) direction of the $z$ axis. All other waves experience total reflection from the structure.
Thus, only waves with a positive spin current are able to transmit through the network, all other waves are perfectly reflected.
This is confirmed by the expressions for energy flux $\langle \vc P_i^{(t)}\rangle_{z}$ and spin current $\langle\ts J_{\mathrm{S}i}^{(t)}\rangle_{zz}$ for the transmitted fields to the left ($i=1$) and right ($i=2$) of the network
\begin{equation}
\begin{gathered}
    \langle\vc P_1^{(t)}\rangle_{z} = \langle\vc P_2^{(t)}\rangle_{z} = \dfrac{c}{8\p} \lrp{\lrv{a_{1+}}^2 - \lrv{a_{2-}}^2},
    \\
    \langle\ts J_{\mathrm{S}1}^{(t)}\rangle_{zz}  = \dfrac{c}{16 \p \omega}\lrp{\lrv{a_{1+}}^2 + \lrv{a_{2-}}^2 -2 \lrv{a_{1-}}^2}
    \\
    \langle\ts J_{\mathrm{S}2}^{(t)}\rangle_{zz} = \dfrac{c}{16 \p \omega}\lrp{\lrv{a_{1+}}^2 + \lrv{a_{2-}}^2 -2 \lrv{a_{2+}}^2}
\end{gathered}
\label{eq:isolator_tr}
\end{equation}
One can see that only amplitudes of the waves with positive spin current (LCP wave traveling along $z$-axis, and RCP wave traveling against $z$-axis) contribute to energy flow.
Additional terms in the expression of the optical spin current are due to the nonzero reflection and do not contribute in the transmitted field.
Thus, such a network is an {\it optical spin isolator}. Moreover, when $\lvert a_{1+} \rvert = \lvert a_{2-} \rvert$, the energy flow through the network is zero, while the optical spin current reaches its maximum value. This regime clearly demonstrates the transfer of optical spin current without accompanying energy transfer.

While the isolator is the simplest non-reciprocal system, more complex systems can be built on its basis. For example, a combination of three isolators connected clockwise will represent an optical circulator (see Figs.~\ref{fig:diode_ideal}d,~\ref{fig:diode_ideal},~e~\ref{fig:diode_ideal}f).
Interestingly, scattering matrix of such devices is both symmetric and hermitian.
It is well-known that only devices that violate Lorentz reciprocity, i.e. have asymmetrical scattering matrix, may be candidates for non-reciprocal optical devices~\cite{jalas2013and}.
However, both isolator and circulator presented in Fig.~\ref{fig:diode_ideal} do not require breaking of the Lorentz reciprocity as energy can be transferred in opposite directions symmetrically.
This is a consequence of the fact that the information transfer in the presented devices is carried out by means of optical spin currents, not electromagnetic energy flow.
However, it is important to mention that the creation of optical nonreciprocity for spin currents still requires the violation of certain symmetries.
The structures and/or materials from which the real-life implementations of the aforementioned devices will be made must have non-zero circular dichroism (CD), since in structures with zero CD the transmission coefficients for LCP and RCP waves are equal, and, as a consequence, the flow of optical spin currents in devices with zero CD will be reciprocal.

\section{Optical spin diode.}
\begin{figure*}[t]
	\centering
	\includegraphics[width=\textwidth]{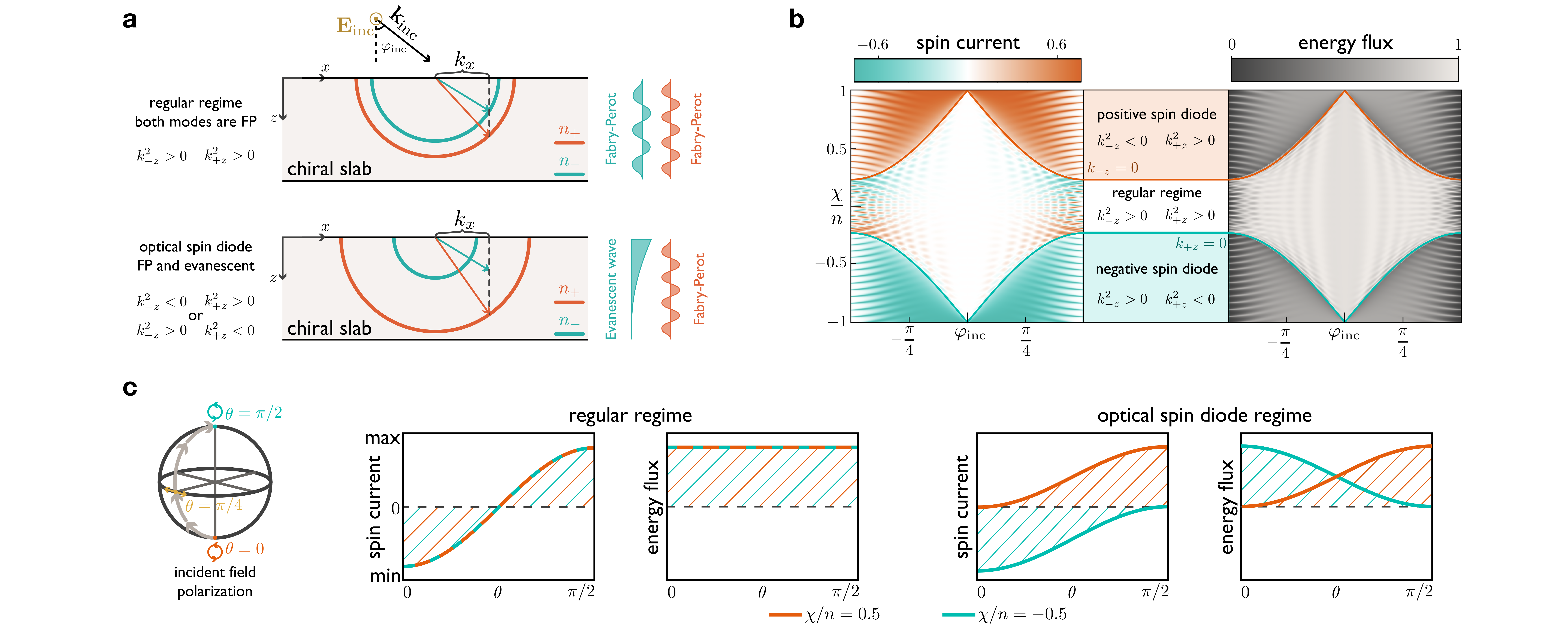}
	\caption{
    Chiral slab as an optcal spin diode. \textbf a – geometry of the scattering problem. \textbf b – optical spin current and energy flux of the transmitted field as functions of angle of incidence $\ph_\mathrm{inc}$ and normalized chirality parameter of the slab $\chi/n$. \textbf c – optical spin current and energy flux of the transmitted field in regular and optical spin diode regimes as a function of the polarization of the incident wave. Inset on the left shows the Poincaré sphere with path along which the polarization was changing.
    }
	\label{fig:diode_chiral}
\end{figure*}

A vast variety of optical structures and devices possess non-zero CD as their intrinsic or engineered property.
Examples of such structures are various optical resonators, metasurfaces, plasmonic structures, and so on \cite{govorov2010theory, govorov2011chiral, hu2017all, kong2018photothermal}.
However, as a simplest example, one can take any optical structure made of chiral materials.
This follows from the fact that eigenmodes of isotropic chiral media with permittivity $\e$, permeability $\m$, and chirality parameter $\chi$ are left (LH) and right (RH) plane waves with 
electric field (see Fig.~\ref{fig:diode_chiral}a and Appendix, Sec.~\ref{apdx:chiral})
\begin{equation}
    \vc E\lrp{\vc r} = \dfrac{a_0}{\sqrt{2}}\lrp{\uv{x} + i \x \s_\a \uv{y}}e^{i \xi n_\a k z},
    \label{eq:field_chiral}
\end{equation}
where $a_0$ is the complex amplitude, $n_\a = \sqrt{\e \m} + \s_\a\chi$ is refractive index for LH ($\a = \ell$, $\s_\ell = 1$) and RH ($\a = \r, \s_\r =-1$) waves, and $\x = \pm 1$ encodes propagation direction.
Note that to simplify the reasoning and improve its clarity we consider here only waves propagating along or against the $z$ axis. The direction of spin current in a plane wave is closely related to its handedness.
For example, for electromagnetic field given by Eq.~\eqref{eq:field_chiral} sign of spin current is determined by the value of $\sigma_\alpha$,
\begin{equation}
    \langle \ts J_\mathrm{S} \rangle_{zz} = \lrv{a_0}^2 n_\a \sigma_\a
\end{equation}
Thus, chiral media is an ideal candidate for creating optical spin devices since different values of the refractive index, and, accordingly, absence of degeneracy in the dispersion of LH and RH waves, results in the difference in the transmission coefficients for waves with positive and negative spin currents.

Even the simplest optical structures made of chiral materials may posses asymmetric transfer characteristics with respect to optical spin currents.
A particular example of such a structure is a slab made of isotropic chiral material (see Fig.~\ref{fig:diode_chiral}a).
Figure~\ref{fig:diode_chiral}b shows optical spin current and energy flux in the transmitted field, as a function of the angle of incidence and normalized chirality parameter $\chi/n$ of the slab, for the case of s-polarized incident wave (see Appendix, Sec.~\ref{apdx:scattering} for details).
Such a polarization was chosen for the sake of clarity, since it contains both LH and RH waves.
Based on the sing and the amplitude of spin current, one can distinguish three different regimes.
First (regular) regime, is characterized by the absence of spin currents in the transmitted field, while in second and third (positive and negative spin diode) regimes, spin current is either strictly positive or strictly negative (see Fig.~\ref{fig:diode_chiral}c).
The key factor for understanding the existence of the aforementioned regimes is the fact that the refractive indices for LH and RH waves are different in chiral media.
Depending on the angle of incidence and material parameters of the slab, the incident electromagnetic wave experience either regular refraction, or frustrated total internal reflection (FTIR).
In the second case, the amplitude of the transmitted field will exponentially decrease as thickness of the slab increases.
Due to the difference in refractive indices, the FTIR critical angle is different for LH and RH waves.
Thus, in $\lrp{\ph_\mathrm{inc}, \chi/n}$ parametric space there exist areas in which LH (RH) wave will be dominant in the transmitted field, while RH (LH) wave will be negligible due to FTIR.
Therefore, only positive (negative) spin currents will be able to flow through the slab.
It is important to note that in the analysis above, only incidence along $z$ axis was considered for the sake of clarity.
However, due to the geometry and chirality of the slab, considered system is invariant under transformation $\lrp{z, \chi}\to\lrp{-z, - \chi}$, which preserves handedness of the waves.
Thus, all presented analysis is valid for waves propagating against $z$ axis.
Which means that similar to freespace and optical spin isolator, such chiral slab also supports field configurations with zero poynting vector and nonzero optical spin current.

\section{Information transfer via optical spin currents}
\begin{figure}[t]
	\center
	\includegraphics[width=0.8\linewidth]{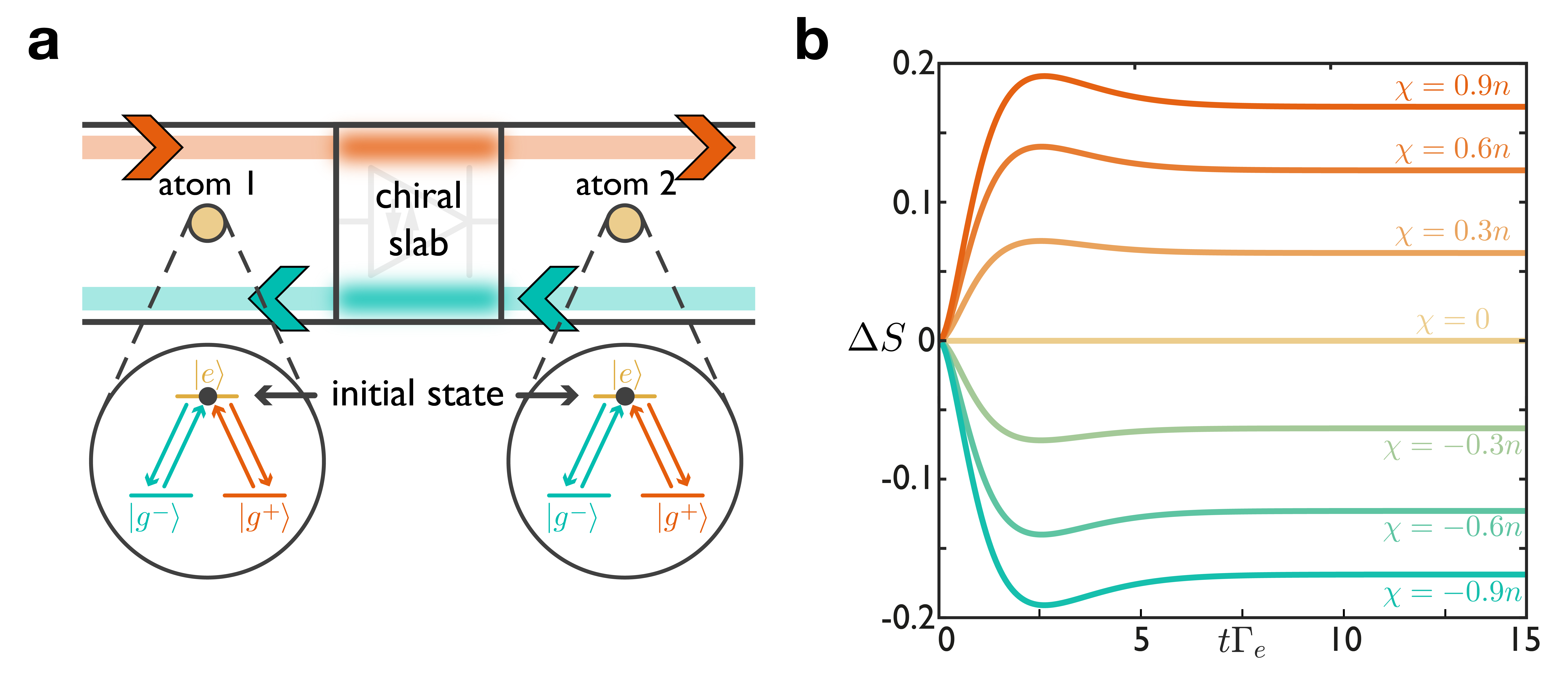}
	\caption{Interaction of two lambda-atoms via chiral optical spin diode. \textbf{a} – geometry of the problem. \textbf b – evolution of spin difference between first and second atoms. Calculations were done for the slab with thickness $d=0.1\lambda$ and permittivity $\e=2$. Distance between each dipole and surface of the slab was taken equal to $d_z=0.05\lambda$.}
	\label{fig:quantum}
\end{figure}
Nonreciprocal transfer is crucial for the creation of optical information processing and transmission systems, since it improves the quality of signal transmission, protecting against backscattering and reflections \cite{zhou2024nonreciprocal, mann2019nonreciprocal}, and enabling advanced optical functions such as optical switches, asymmetric power limiters, etc~\cite{wang2025self, tripathi2024nanoscale}.
Optical spintronics not only offers an efficient way to create nonreciprocal devices, since creating asymmetric spin currents does not require breaking Lorentz reciprocity, but is also a promising way to create energy-efficient photonic systems, since optical spin transfer does not require electromagnetic energy transfer.
As an example of information transfer via optical spin currents between two systems, we consider two lambda-atoms interacting via chiral optical spin diode considered in the previous section (see Fig.~\ref{fig:quantum}a).
The energy levels of each atom are chosen in such a way that each transition interacts only with LCP or RCP polarized wave,
\begin{equation}
    \bra{g^\pm} \hat{\vc d} \ket{e} = \mp \dfrac{d_0}{\sqrt{2}}\lrp{\uv{x} \pm i \uv{y}},
\end{equation}
where $\hat{\vc d}$ is the transition dipole moment operator. 
We prepare both atoms initially in the excited state and calculate the evolution of the system using the master equation for the density matrix of atomic system.
For details on the calculations, refer to Sec.~\ref{apdx:supl_quantum}.
Nonreciprocal nature of optical spin current flow in optical spin diode will leave an imprint on the interaction between two atoms.
Because LCP (RCP) photons emitted by the first atom have higher a probability to be transmitted through the diode from left to right, and RCP (LCP) photons emitted by the second atom have a higher probability to be transmitted through the diode from right to left for positive (negative) chirality parameter, the interaction between $\ket{g^+}$ and $\ket{g^-}$ ground states of different atoms will be asymmetrical.
Therefore, first atom will have a higher probability to fall in $\ket{g^-}$ ($\ket{g^+}$) and second atom will have a higher probability to fall in $\ket{g^+}$ ($\ket{g^-}$) state as $t\to \infty$ for positive (negative) chirality parameter.
If one consider each of the state to correspond to bit 0 or bit 1 states, then aforementioned behavior will mean that information was transferred from one atom to another.
Figure \ref{fig:quantum}b shows temporal evolution of difference between spin z-projections of the first and second atom, defined as
\begin{equation}
    \Delta S_{z} = \langle \hat S_{z,2} \rangle - \langle \hat S_{z, 1} \rangle,
\end{equation}
where $\hat S_{z,i}$ is the spin operator for the $i$'th atomic subsystem, for different values of the chirality parameter.
It can be seen that over time the spin of the system takes on a non-zero value, the amplitude of which at $t\to\infty$ depends on the chirality parameter $\chi$. Therefore the larger the chirality parameter, the higher the stationary value of the spin. 

\section{Conclusions}
We have shown that optical spin transfer can occur independently of electromagnetic energy transfer. For example, a superposition of LCP and RCP waves propagating in opposite directions will have zero energy flux but non-zero optical spin current. Moreover, by controlling the material properties of the systems, it is possible to achieve non-reciprocal transfer characteristics with respect to spin currents, which opens the way to the creation of non-reciprocal optical spin devices, such as an optical spin diode and an optical spin isolator. Interestingly, the creation of such devices is possible without violating Lorentz reciprocity, due to which the above-mentioned systems will be non-reciprocal for spin currents but reciprocal in terms of electromagnetic energy. Real-life implementations of such devices can be performed based on optical systems possessing non-zero circular dichroism, such as, for example, chiral media and metamaterials. Since the creation of nonreciprocal optical spin transfer does not require violation of Lorentz reciprocity and hermicity in the system, the creation of nonreciprocal optical spin devices is much simplified compared to regular optical diodes, circulators, and similar devices.

Moreover, the fact that transfer of the optical spin do not require transfer of electromagnetic energy makes encoding information with optical spin a promising path towards energy-efficient photonics.

\begin{backmatter}

\bmsection{Acknowledgments}
The studies were supported by the Russian Science Foundation (Project 23-72-10059). I.D. acknowledges the BASIS foundation for valuable support.

\end{backmatter}

\appendix
\section{Continuity equation for SAM}
\label{apdx:SAM_continuity}
We follow \cite{bliokh2013dual} for the definition of SAM for arbitrary real-valued fields
\begin{equation}
    \vc S = \dfrac{1}{8\p} \lrb{ \vc E \times \vc A + \vc B \times \vc C},
\end{equation}
where $\vc C$ and $\vc A$ are electric and magnetic vector potentials defined as
\begin{equation}
    \vc E = -\rot \vc C, \quad \vc B = \rot \vc A.
\end{equation}
As in \cite{bliokh2013dual}, we work in the Coulomb gauge
\begin{equation}
    \div \vc C = 0, \quad \div \vc A = 0,
\end{equation}
with both scalar potentials equal to zero.
To derive the continuity equation we start from time derivative of $\vc S$,
\begin{equation}
    \dfrac{1}{8\p}\dfrac{\partial}{\partial t} \lrb{\vc E \times \vc A + \vc B \times \vc C} = \dfrac{1}{8\p}\dfrac{\partial \vc E}{\partial t} \times \vc A + \dfrac{1}{8\p}\vc E \times \dfrac{\partial \vc A}{\partial t} +\dfrac{1}{8\p}\dfrac{\partial \vc B}{\partial t}\times \vc C + \dfrac{1}{8\p}\vc B \times \dfrac{\partial \vc C}{\partial t}.
    \label{eq:s:dsdt}
\end{equation}
From Maxwell equations
\begin{equation}
    \dfrac{\partial \vc E}{\partial t} = c\rot \vc B, \quad \dfrac{\partial \vc B}{\partial t} = - c \rot \vc E,
    \label{eq:s:sub1}
\end{equation}
and due to the Coulomb gauge
\begin{equation}
    \dfrac{\partial \vc C}{\partial t} = - \vc B, \quad \dfrac{\partial \vc A}{\partial t} = - \vc E.
    \label{eq:s:sub2}
\end{equation}
Substitution of Eqs.~\eqref{eq:s:sub1}, \eqref{eq:s:sub2} into Eq.~\eqref{eq:s:dsdt} gives
\begin{equation}
   \dfrac{1}{8\p}\dfrac{\partial}{\partial t} \lrb{\vc E \times \vc A + \vc B \times \vc C} = \dfrac{c}{8\p}\lrp{\rot \vc B}\times \vc A - \dfrac{c}{2}\lrp{\rot \vc E}\times \vc C
   \label{eq:s:dsdt2}
\end{equation}
We assume that RHS of Eq.~\eqref{eq:s:dsdt2} can be expressed as
\begin{equation}
    \dfrac{c}{8\p}\lrp{\rot \vc B}\times \vc A - \dfrac{c}{8\p}\lrp{\rot \vc E}\times \vc C = - \div \ts J_{\mathrm S},
\end{equation}
so that
\begin{equation}
    \dfrac{\partial \vc S}{\partial t} + \div \ts J_\mathrm{S} = 0,
\end{equation}
where $\ts J_{\mathrm S}$ is an optical SAM flux density (i.e. optical spin current).
We now consider monochromatic fields
\begin{equation}
    \vc F = \dfrac{1}{2}\Big(\vc F\lrp{\vc r}e^{- i \o t} + \vc F^*\lrp{\vc r}e^{i \o t}\Big).
\end{equation}
We are interested in the time-averaged value of the optical spin current,
which we calculate via RHS of Eq.~\eqref{eq:s:dsdt2} as
\begin{equation}
   \lra{\mathrm{R.H.S}} = -\div \langle \ts J_\mathrm{S} \rangle = -\lra{\div \ts J_\mathrm{S}}.
\end{equation}
For monochromatic fields
\begin{multline}
    \lra{\mathrm{R.H.S}} = 
    \\
    =
    \dfrac{c}{32 \p T} \int\limits_0^T \Big[\lrp{\rot \lrb{\vc B\lrp{\vc r} e^{- i \o t} + \vc B^*\lrp{\vc r} e^{i \o t}}} \cdot \lrp{\vc A\lrp{\vc r} e^{- i \o t} + \vc A^*\lrp{\vc r} e^{i \o t}}\Big] \mathrm{d} t  - \text{E.P.} = 
    \\
    = \dfrac{c}{16 \p}\Re\lrc{\lrp{\rot \vc B\lrp{\vc r}}\times \vc A^{*}\lrp{\vc r}} - \text{E.P.},
\end{multline}
where "E.P." encodes electric part of the expression, and $T = 2 \p /\o$.
We then use Eq.~\eqref{eq:s:sub2} for monochromatic fields to write
\begin{multline}
    \lra{\mathrm{R.H.S}} = \dfrac{i c}{16 \p 
    \o}\Re\Big\{ \lrp{\rot \vc B\lrp{\vc r} }\times \vc E^*\lrp{\vc r} - \lrp{\rot \vc E\lrp{\vc r} }\times \vc B^*\lrp{\vc r}\Big\} = 
    \\
   = -\dfrac{c}{16 \p \o}\Im\Big\{\vc B^*\lrp{\vc r} \times \lrp{\rot \vc E\lrp{\vc r} } +\vc E\lrp{\vc r} \times \lrp{\rot \vc B^*\lrp{\vc r}}\Big\}.
\end{multline}
We then use following identity
\begin{equation}
    \nabla(\mathbf{A} \cdot \mathbf{B})=(\mathbf{A} \cdot \nabla) \mathbf{B}+(\mathbf{B} \cdot \nabla) \mathbf{A}+ \mathbf{A} \times(\nabla \times \mathbf{B})+\mathbf{B} \times(\nabla \times \mathbf{A}),
\end{equation}
and note that
\begin{gather}
    \nabla\lrp{\vc A \cdot \vc B} = \div\lrp{\tsone\: \vc A \cdot \vc B}, 
    \nonumber\\ 
    \lrp{\vc A\cdot\nabla}\vc B = \div\lrp{\vc A\otimes \vc B} - \vc A\lrp{\div \vc B}.
\end{gather}
Finally, we arrive at
\begin{equation}
    \lra{\mathrm{R.H.S.}} =  -\dfrac{c}{16 \p \o} \cdot 
 \div \Bigg[
        \Im \Big\{
            \tsone\lrp{\vc E\lrp{\vc r} \cdot \vc B^*\lrp{\vc r}} -\vc E\lrp{\vc r}\otimes \vc B^*\lrp{\vc r} - \vc B^*\lrp{\vc r}\otimes \vc E\lrp{\vc r}
        \Big\}
    \Bigg],
\end{equation}
thus concluding that
\begin{equation}
    \langle \ts J_\mathrm{S} \rangle = \dfrac{c}{16 \p \o} \Im \Big\{
            \tsone\lrp{\vc E\lrp{\vc r} \cdot \vc B^*\lrp{\vc r}} -\vc E\lrp{\vc r}\otimes \vc B^*\lrp{\vc r} - \vc B^*\lrp{\vc r}\otimes \vc E\lrp{\vc r}
        \Big\}.
\end{equation}

\section{Energy, Poynting vector, SAM, and optical spin current for plane wave and wavepacket}
\label{apdx:wavepacket}
We consider arbitrary plane wave in vacuum with electric and magnetic fields given by
\begin{equation}
\begin{aligned}
    &\vc E\lrp{\vc r} = \lrp{\uv{1}a_1 + \uv{2}a_2}e^{i \vc k \cdot \vc r},
    \\
    &\vc B\lrp{\vc r} = \lrp{-\uv{1}a_2 + \uv{2}a_1}e^{i \vc k \cdot \vc r},
    \end{aligned}
\end{equation}
where $a_1$, $a_2$ are complex amplitudes of two orthogonal linear polarizations,
and $\uv{1}$, $\uv{2}$ are real-valued unit vectors given by
\begin{equation}
    \uv{1}\times \uv{2} = \uv{k}, \quad \uv{k} = \vc k/k, \quad \uv{i}\cdot \uv{j} = \d_{ij}.
\end{equation}
We calculate time-averaged energy density, Poynting vector, SAM density, and optical spin current density as follows~\cite{bliokh2017optical, jackson1999classical}
\begin{equation}
    \lra{W} = \dfrac{1}{16 \p} \lrp{\lrv{\vc E^*\lrp{\vc r}}^2 + \lrv{\vc B\lrp{\vc r}}^2 } = \dfrac{1}{8 \p}\lrp{\lrv{a_1}^2 + \lrv{a_2}^2},
\end{equation}
\begin{equation}
    \lra{\vc P} = \dfrac{c}{8 \p} \Re\lrc{\vc E\lrp{\vc r}\times \vc B^*\lrp{\vc r} } = \uv{k}\dfrac{c}{8\p}\lrp{\lrv{a_1}^2 + \lrv{a_2}^2},
\end{equation}
\begin{equation}
    \lra{\vc S} = \dfrac{1}{16 \p \o}\Im\lrc{\vc E^*\lrp{\vc r}\times \vc E\lrp{\vc r} + \vc B^*\lrp{\vc r}\times \vc B\lrp{\vc r}} = \\ \uv{k} \dfrac{1}{4 \p \o} \Im\lrc{a_1^* a_2},
\end{equation}
\begin{multline}
    \langle \ts J_\mathrm{S} \rangle =  \dfrac{c}{16 \p \o}\Im\big\{\tsone \lrp{\vc E\lrp{\vc r}\cdot \vc B^*\lrp{\vc r}} -\vc E\lrp{\vc r}\otimes \vc B^*\lrp{\vc r} - \vc B^*\lrp{\vc r}\otimes \vc E\lrp{\vc r}\big\} = 
    \\
    = \uv{k}\otimes \uv{k} \dfrac{c}{8 \p \o}\Im\lrc{a_1^* a_2}.
\end{multline}
Thus, one can conclude that for a plane wave, optical spin current density can be calculated as
\begin{equation}
    \langle \ts J_\mathrm{S} \rangle = \dfrac{1}{2}\lra{\vc S}\otimes \vc v_\mathrm{g},
\end{equation}
where
\begin{equation}
    \vc v_\mathrm{g} = \dfrac{ \lra{\vc P}}{\lra{W}} = c\: \uv{k}
\end{equation}
is the group velocity.

From the orthogonality of plane waves with different wavevectors, it is easy to show that optical spin current for a wavepacket will be equal to a sum of current of each plane wave.
I.e. for a wavepacket with electric and magnetic fields given by
\begin{equation}
\begin{aligned}
    &\vc E\lrp{\vc r} = \int\lrp{\uv{1}a_1\lrp{\vc k} + \uv{2}a_2\lrp{\vc k}}e^{i \vc k \cdot \vc r} \mathrm{d}^3 k,
    \\
    &\vc B\lrp{\vc r} = \int\lrp{-\uv{1}a_2\lrp{\vc k} + \uv{2}a_1\lrp{\vc k}}e^{i \vc k \cdot \vc r} \mathrm{d}^3 k,
    \end{aligned}
\end{equation}
total optical spin current can be calculated as
\begin{equation}
    \lra{\lra{\ts J_s}} = \dfrac{c}{8 \p \o}\int \Im\lrc{ a_1^*\lrp{\vc k} a_2\lrp{\vc k}}\mathrm{d}^3 k,
\end{equation}
where double brackets encode both spatial integration and temporal averaging.

\section{Eigenmodes of Isotropic Chiral Media}
\label{apdx:chiral}
We start from Maxwell equations for monochromatic fields in chiral media described by permittivity $\e$, permeability $\m$, and chirality parameter $\chi$,
\begin{equation}
    \begin{gathered}
        \rot \vc E\lrp{\vc r} = i k \vc B \lrp{\vc r}, \quad \rot \vc H\lrp{\vc r} = - i k \vc D\lrp{\vc r},
        \\
        \div \vc D\lrp{\vc r} = 0, \quad \div \vc B\lrp{\vc r} = 0,
    \end{gathered}
    \label{eq:s:me}
\end{equation}
where
\begin{equation}
\begin{gathered}
    \vc D\lrp{\vc r} = \e \vc E\lrp{\vc r} + i \chi \vc H\lrp{\vc r}, \quad \vc B\lrp{\vc r} = \m \vc H\lrp{\vc r} - i \chi \vc E\lrp{\vc r}.
    \end{gathered}
    \label{eq:s:mm}
\end{equation}
Due to isotropicity and translational symmetry of media in all directions, it is possible to assume plane-wave solutions, and fix $z$ as propagation direction, i.e.
\begin{equation}
    \vc E\lrp{\vc r} = \vc E_0 e^{i \b k z}, \quad \vc H\lrp{\vc r} = \vc H_0 e^{i \b k z},
    \label{eq:s:pw}
\end{equation}
where $k = \o/c$ is the wavevector, $\b$ is a dimensionless propagation constant, and $\vc E_0$ and $\vc H_0$ are complex amplitudes of electric and magnetic fields respectively.
We substitute Eq.~\eqref{eq:s:pw} into Eqs.~\eqref{eq:s:me},~\eqref{eq:s:mm} and arrive at
\begin{equation}
        i \b \ts e_\times \vc E_0 = i \m \vc H_0 + \chi \vc E_0, \quad i \b \ts e_\times \vc H_0 = - i \e \vc E_0 + \chi \vc H_0.
\end{equation}
where
\begin{equation}
    \ts e_\times = 
    \begin{bmatrix}
        0 & - 1 & 0
        \\
        1 & 0 & 0
        \\
        0 & 0 & 0
    \end{bmatrix}.
\end{equation}
We express electric field via magnetic and vice-versa
\begin{equation}
        \vc E_0 = \dfrac{1}{- i\e}\lrb{i \b \ts e_\times - \chi}\vc H_0, \quad \vc H_0 = \dfrac{1}{i \m}\lrb{i\b\ts e_\times - \chi}\vc E_0,
\end{equation}
thus arriving at the equation
\begin{equation}
    \lrb{\lrp{i \b \ts e_\times - \chi }^2 - \e \m} \vc E_0 = 0,
\end{equation}
which solution is
\begin{equation}
    \vc E_0 = a_0 \lrb{1,i \s \x ,0}, \quad n_\s = \sqrt{\e \m} + \s\chi,
\end{equation}
where $\s = \pm 1$ encodes handedness of the wave, and $\x=\pm 1$ encodes propagation direction.
Finally, electric and magnetic fields are
\begin{equation}
        \vc E\lrp{\vc r} = a_0 \lrb{1 ,i \s \x, 0} e^{i \x n_\s k z}, \: \: \vc H\lrp{\vc r} = a_0 \lrb{-i \s \x, 1, 0} e^{i \x n_\s k z}.
\end{equation}

\section{Scattering Matrix for Isotropic Chiral Slab}
\label{apdx:scattering}
We follow the definition of incident and scattering channels similar to Eq.~\eqref{eq:sc_field}, but consider linear polarization and both normal and oblique incidence.
Thus, the electric field is given by the following expression:
\begin{equation}
    \vc E_i\lrp{\vc r} = \lrp{\uv{s}a_{is} + \uv{ip}^aa_{ip}} e^{i k_x x + i \b_i k_z z + i k_z a} +
    \lrp{\uv{s}b_{is} + \uv{ip}^bb_{ip}} e^{i k_x x - i \b_i k_z - i k_z a},
\end{equation}
where
\begin{equation}
    \uv{s} = \uv{y},\quad 
    \uv{ip}^a = -\uv{x}\dfrac{\b_i k_z}{k} + \uv{z}\dfrac{k_x}{k},\quad 
    \uv{ip}^b = \uv{x}\dfrac{\b_i k_z}{k} + \uv{z}\dfrac{k_x}{k},
\end{equation}
$k_z = \sqrt{k^2 - k_x^2}$, and $a$ is the half-thickness of the slab.
The scattering matrix for chiral slab $\ts{\mathrm S}$ can be obtained by solving system of boundary equations for both boundaries of the slab and writes as
\begin{equation}
	\ts S = \ts M \cdot \begin{bmatrix}
	    \ts \Scal_\mathrm{odd} & \ts 0
     \\
     \ts 0 & \ts \Scal_\mathrm{even}
	\end{bmatrix} \cdot \ts M^{-1},
\end{equation}

where $\ts 0$ is 2 by 2 matrix with zeros,
matrices $\ts \Scal_\mathrm{odd}$ and $\ts \Scal_\mathrm{even}$ are
\begin{align}
        \ts \Scal_\mathrm{odd} & = 
        \begin{bmatrix}
            \dfrac{t_{-\r} t_{+\ell} - \k_-^2}{t_{-\r} t_{-\ell} - \k_-^2} & \dfrac{\lrp{t_{+\r} - t_{-\r} } \k_-}{t_{-\r} t_{-\ell} - \k_-^2}
	\vspace{4pt}\\
	\dfrac{\lrp{t_{+\ell} - t_{-\ell} } \k_-}{t_{-\r} t_{-\ell} - \k_-^2} & \dfrac{t_{-\ell} t_{+\r} - \k_-^2}{t_{-\r} t_{-\ell} - \k_-^2}
        \end{bmatrix}
        , \nonumber\\ \ts \Scal_\mathrm{even} & = \begin{bmatrix}\dfrac{c_{-\ell} c_{+\r} - \k_-^2}{c_{+\r} c_{+\ell} - \k_-^2} & \dfrac{\lrp{c_{-\r} - c_{+\r} } \k_-}{c_{+\r} c_{+\ell} - \k_-^2}
	\vspace{4pt}\\
	\dfrac{\lrp{c_{-\ell} - c_{+\ell} } \k_-}{c_{+\r} c_{+\ell} - \k_-^2} & \dfrac{c_{-\r} c_{+\ell} - \k_-^2}{c_{+\r} c_{+\ell} - \k_-^2},
 \end{bmatrix}
\end{align}
with matrix elements defined as
\begin{gather}
	t_{\pm, \a} = \k_+ \pm 2 i \dfrac{k_{\a x}}{k_\a} \tan\lrp{a k_{\a x}}, \nonumber\\
    c_{\pm, \a} = \k_+ \pm 2 i \dfrac{k_{\a x}}{k_\a} \cot\lrp{a k_{\a x}},
\end{gather}
where $\k_\pm = k_y/k_0 \lrp{n^{-1} \pm n}$, $k_{\a x} = \sqrt{k_\a^2 - k_y^2}$, $k_\a = n_\a k_0$, with index $\a = \ell, \r$, refractive index $n_\ell = \sqrt{\e \m} + \s_\a\chi$, $\s_\ell = 1$, $\s_\r = -1$, 
and matrix $\ts M$ defined as
\begin{equation}
	\ts M = \dfrac{1}{n}\begin{bmatrix}
		n & n & n & n
		\\
		-i & i & -i  & i
		\\
		n & n & -n & -n
		\\
		-i & i & i & -i
	\end{bmatrix}.
\end{equation}

\section{Theoretical framework for the spin transfer between two $\lambda$-atoms}
\label{apdx:supl_quantum}
\subsection{Master equation}
For a set of two-level atoms in an arbitrary optical environment we can write the following master equation for the density matrix\cite{GangarajPRA2017}:
\begin{gather}
    \dot{\hat \rho}(t) = - \frac{i}{\hbar} \left[  {\hat H}_s, \hat \rho(t) \right] + \mathcal{\hat L} \hat \rho(t), \nonumber\\
\mathcal{\hat L} \hat \rho(t) = \sum\limits_{i} \frac{\Gamma_{ii}}{2} \left( 2 \hat \sigma_i^- \hat \rho(t) \hat \sigma_i^+ - \hat \sigma_i^{ee} \hat \rho(t) - \hat \rho(t) \hat \sigma_i^{ee} \right) + \\
\sum\limits_{{}^{i,j}_{i \ne j} }  \left( - i G_{ij} \left[ \hat \sigma_j^- \hat \rho(t), \hat \sigma_i^+ \right]  + i G_{ij}^* \left[ \hat \sigma_i^-, \hat \rho(t) \hat \sigma^+_j \right] \right),
\end{gather}
where indeces $i,j$ run through atoms, $\hat H_s = \sum\limits_i \hbar \Delta \omega_i \hat \sigma_i^{ee}$ accounts for Lamb shifts for each atom, $\Delta \omega_i = \omega_{0,i} - g_{ii}$, $\hat \sigma^-_i = |g_i\rangle \langle e_i|$, $\hat \sigma^{ee}_i = |e_i \rangle \langle e_i |$ are spin operators, $G_{i,j} = 4 \pi k_0^2 \mathbf{d}_i^\dagger \mathbf{G}(\mathbf{r_i}, \mathbf{r_j}, \omega_0) \mathbf{d}_j/\hbar$ is a complex-valued coupling constant of atoms $i, j$ that is defined by the classical electromagnetic Green's function of the environment $\mathbf{G}(\mathbf{r_i}, \mathbf{r_j}, \omega_0)$, $k_0 = \omega_0/c = 2 \pi/\lambda_0$ is the resonant wavenumber, $\mathbf{r_i}$ are locations of atoms, $g_{ii} = \text{Re} \: G_{i,i}, \Gamma_{ii} = 2\: \text{Im} \: G_{i,i}$.

For $\lambda$-atoms with two degenerate Zeeman ground state sublevels $| g_- \rangle, | g_+ \rangle$, and a single excited state $| e \rangle$, by treating each transition in each atom as an individual dissipator one can write a similar set of equations:
\begin{multline}
    \mathcal{\hat L} \hat \rho(t) =
    \sum\limits_{k}  \dfrac{\Gamma_{k, k}^{(+,+)}}{2} \left[ 2 \hat \sigma_{g_{k}^+, e_k} \hat \rho(t) \hat \sigma_{e_k, k+} - \hat \sigma_{e_k, e_k} \hat \rho(t) - \hat \rho(t) \hat \sigma_{e_k, e_k} \right]  + \\ +
    \sum\limits_{k} \dfrac{\Gamma_{k, k}^{(-,-)}}{2} \left[ 2 \hat \sigma_{g_k^-, e_k} \hat \rho(t) \hat \sigma_{e_k, g_k^-} - \hat \sigma_{e_k, e_k} \hat \rho(t) - \hat \rho(t) \hat \sigma_{e_k, e_k} \right] + \\ +
    \sum\limits_{k,l} \left[ - i G_{k, l}^{(+,-)} \left( \hat \sigma_{g^-_l, e_l} \hat \rho(t)  \hat \sigma_{e_k, g^+_k} - \hat \sigma_{e_k, g^+_k} \hat \sigma_{g^-_l, e_l} \hat \rho(t) \right)  + h.c.  \right] + \\ +
    \sum\limits_{k,l} \left[ - i G_{k, l}^{(-,+)} \left( \hat \sigma_{g^+_l, e_l} \hat \rho(t) \hat \sigma_{e_k, g^-_k} - \hat \sigma_{e_k, g^-_k} \hat \sigma_{g^+_l, e_l} \hat \rho(t) \right) + h.c.  \right] + \\ +
    \sum\limits_{{}^{k,l}_{k \ne l}} \left[  - i G_{k, l}^{(+,+)} \left( \hat \sigma_{g_l^+, e_l} \hat \rho(t) \hat \sigma_{e_k, g^+_k} - \hat \sigma_{e_k, g^+_k} \hat \sigma_{g^+_l, e_l} \hat \rho(t) \right)  + h.c. \right] + \\ +
    \sum\limits_{{}^{k,l}_{k \ne l}} \left[ - i G_{k, l}^{(-,-)} \left( \hat \sigma_{g_l^-, e_l} \hat \rho(t) \hat \sigma_{e_k, g^-_k} - \hat \sigma_{e_k, -k} \hat \sigma_{g^-_l, e_l} \hat \rho(t) \right) + h.c. \right],
\end{multline}
where $G_{k,k'}^{(\sigma, \sigma')} = 4 \pi k_0^2 \mathbf{d}^{\sigma \dagger}_{k} \mathbf{G}(\mathbf{r}_k, \mathbf{r}_{k'}, \omega_0) \mathbf{d}_{k'}^{\sigma'}/\hbar$.
\subsection{Analytical result}
The above system of $81$ equations for two atoms in the most general case can be solved numerically. However, we can also obtain some analytical results, for instance, when we consider two identical emitters, which means they have equal emission rates for both transitions, and equal Lamb shifts ($\Gamma_{1, 1}^{(+,+)} = \Gamma_{1, 1}^{(-,-)} = \Gamma_{2, 2}^{(+,+)} = \Gamma_{2, 2}^{(-,-)} = \Gamma$, $g_{11} = g_{22}$), and also we will consider a perfectly chiral coupling ($G_{2,1}^{(-,-)} = G_{1,2}^{(+,+)}$, while all other coupling constants are zero). In order to shorten notations let us rename states in the following way $1 \to g_1^- g_2^-, 2 \to g_1^- g_2^+, 3 \to g_1^- e_2, 4 \to g_1^+ g_2^-, 5 \to g_1^+ g_2^+, 6 \to g_1^+ e_2, 7 \to e_1 g_2^-, 8 \to e_1 g_2^+, 9 \to e_1 e_2$, and then we can write the system of equations on the density matrix components:
\begin{gather}
    \dot{\rho}_{22}(t) = \Gamma \rho_{33}(t) + \Gamma \rho_{88}(t), \nonumber\\
    \dot{\rho}_{33}(t) = \Gamma \rho_{99}(t) - 2 \Gamma \rho_{33}(t) + i G \rho_{73}(t) - i G^* \rho_{37}(t), \nonumber\\
    \dot{\rho}_{73}(t) = - 2 \Gamma \rho_{73}(t) + i G^* \rho_{99}(t) - i G^* \rho_{77}(t), \nonumber\\
    \dot{\rho}_{37}(t) = - 2  \Gamma \rho_{37}(t) - i G \rho_{99}(t) + i G \rho_{77}(t), \nonumber\\
    \dot{\rho}_{88}(t) = - 2 \Gamma \rho_{88}(t) + \Gamma \rho_{99}(t) + i G \rho_{6 8}(t) - i G^* \rho_{8 6}(t), \nonumber\\
    \dot{\rho}_{7 7}(t) = - 2 \Gamma \rho_{77}(t) + \Gamma \rho_{99}(t), \nonumber\\
    \dot{\rho}_{6 8}(t) = - 2 \Gamma \rho_{6 8}(t) + i G^* \rho_{9 9}(t) - i G^* \rho_{6 6}(t), \nonumber\\
    \dot{\rho}_{8 6}(t) = - 2 \Gamma \rho_{8 6}(t) - i G \rho_{9 9}(t) + i G \rho_{6 6}(t), \nonumber\\
    \dot{\rho}_{6 6}(t) = \Gamma \rho_{9 9}(t) - 2 \Gamma \rho_{6 6}(t).
    \label{DensMatSys1}
\end{gather}
Let us begin with the following initial condition - atom $1$ is excited, while $2$ is in the $g_2^-$ state, and as there is only a single excitation in the system initially, and no external driving, then we can assume $\rho_{99}(t)=0$. In this case we can obtain the solution by consecutively solving the above system:
\begin{gather}
\rho_{77}(t) = e^{- 2 \Gamma t}, \quad \rho_{73}(t) = - i G^* t e^{- 2 \Gamma t}, \nonumber\\
\rho_{33}(t) = |G|^2 t^2 e^{- 2 \Gamma t}, \nonumber\\
\rho_{22}(t) = \dfrac{|G|^2}{4 \Gamma^2} \left( 1 - e^{- 2 \Gamma t} - 2 \Gamma t \left( 1 + \Gamma t \right) e^{- 2 \Gamma t} \right).
\end{gather}
Let us discuss the obtained result, namely, the component $\rho_{22}(t)$ that is responsible for the spin flip of the second atom. For a general nanophotonic structure it is hard to estimate what is the limit for the relation between $G$, and $\Gamma$ as there are many types of modes involved. For instance, the near fields (or evanescent modes) do not propagate in the far field, and do not contribute to $\Gamma$ in a lossless structure, but they do contribute to $G$. However, in some very specific types of structures we can give some good atguments on that limit. For instance, in the case of a waveguide with a single "chiral" guided mode prosent, when two atoms are placed sufficiently far away from each other, the interaction $G$ only consists of this guided mode contribution. Therefore, in the best case it can be that $\Gamma = |G|$, when the coupling to the guided mode is much larger than to any other. Then for the stationary state we have $\rho_{22}(t \to \infty) = \dfrac{|G|^2}{4 \Gamma^2} = 0.25$.  It basically means that for a waveguide structure only a quarter of the second atom's ground state spin can be flipped in an ideal case. This number can be easily understood with the following rationale. As we have time reversal symmetry, $\Gamma^{(+,+)}_1 = \Gamma^{(-,-)}_2 = \Gamma$, therefore, there is a 50\% chance for the first atom to go to the ground state $|g_1^- \rangle\rangle$ emitting the $\sigma^+$ polarized photon propagating in the direction of the second atom, that is later absorbed by it making the second atom to undergo a transition $|g_2^-\rangle |e_2\rangle$. Now, there is, again, 50\% chance now for the second to make a transition $|e_2\rangle \to |g_2^+\rangle$ that corresponds to the flip of the second atom's ground state spin. Therefore, in total we have 25\% probability of the spin flip.

For the case when two atoms are both excited initially (which is considered in the main text), one can solve the system in a similar manner. In order to find probabilities for both atoms to be in all combinations of ground states, one needs to add to the system \ref{DensMatSys1} the following equations:
\begin{gather}
    \dot{\rho}_{44} = \Gamma \rho_{77} + \Gamma \rho_{66}, \nonumber\\
    \dot{\rho}_{11} = \Gamma \rho_{77} + \Gamma \rho_{33} - i G \rho_{73} + i G^* \rho_{37}, \nonumber\\
    \dot{\rho}_{55} = \Gamma \rho_{88} + \Gamma \rho_{66} - i G \rho_{68} + i G^* \rho_{86}.
\end{gather}
The solution for ground state is then given by:
\begin{gather}
    \rho_{11} = \dfrac{e^{-4 \Gamma t}}{16 \Gamma^2} \Bigg( 4 \Gamma^2 \left( e^{2 \Gamma t} - 1 \right)^2 + \left| G \right|^2 \times 
    \quad \bigg( 9 + e^{4 \Gamma t} - 2 e^{2 \Gamma t} \big( 5 + 2 \Gamma t (\Gamma t - 4) \big) \bigg)  \Bigg), \nonumber
    \\ 
    \rho_{22} = \dfrac{e^{- 4 \Gamma t}}{8 \Gamma^2} \Bigg( 2 \Gamma^2 \left( e^{2 \Gamma t} - 1 \right)^2 - \left| G \right|^2 \times
    \quad \bigg( -3 + e^{4 \Gamma t} + e^{2 \Gamma t} \big( 2 + 4 \Gamma t (\Gamma t - 2) \big)  \bigg) \Bigg), \nonumber
    \\
    \rho_{44} = e^{- 2 \Gamma t} \sinh^2 \left( \Gamma t \right), \quad \rho_{11} = \rho_{55}.
\end{gather}
With this, we can analyze what is the spin polarization of atoms after the relaxation ($t \to \infty$):
\begin{gather}
    S_{1,z}^{(\infty)} = -1 ( \rho_{11}^{(\infty)} + \rho_{22}^{(\infty)} ) + 1 ( \rho_{44}^{(\infty)} + \rho_{55}^{(\infty)} ) = \dfrac{\left| G \right|^2}{8 \Gamma^2}, \nonumber\\
    S_{2,z}^{(\infty)} = - S_{1,z}^{(\infty)}.
\end{gather}
One can see that the resulting spin polarization of a given atom, again, depends on the relation between the coupling constant $G$, and decay rate $\Gamma$. Also, due to the symmetry $1 \leftrightarrow 2, \sigma^- \leftrightarrow \sigma^+$, the polarization of two atoms is equal in magnitude, but opposite in sign.

\bibliography{bibl}

\end{document}